\documentclass[aps,prd,showpacs,twocolumn,superscriptaddress,nofootinbib,preprintnumbers]{revtex4-1} 

\makeatletter
\def\p@subsection{}
\makeatother

\usepackage{color,graphicx,amsmath,amssymb,lipsum}
\usepackage[colorlinks=true,citecolor=blue,linkcolor=blue,urlcolor=blue, backref=false,pdfborder={0 0 0}]{hyperref}

\usepackage[normalem]{ulem}
\usepackage[utf8]{inputenc}

\usepackage{ulem}

\newcommand{\be}{\begin{equation}}
\newcommand{\ee}{\end{equation}}
\newcommand{\beqa}{\begin{eqnarray}}
\newcommand{\eeqa}{\end{eqnarray}}

\renewcommand\k{\bf k}

\renewcommand\t{\theta}

\renewcommand\L{\Lambda}

\def\e{{\rm e}}

\newcommand{\bseq}{\begin{subequations}}
\newcommand{\eseq}{\end{subequations}}

\renewcommand{\ln}{\mathop{\rm ln}\nolimits}

\def\gsim{\raise0.3ex\hbox{$\;>$\kern-0.75em\raise-1.1ex\hbox{$\sim\;$}}}
\def\lsim{\raise0.3ex\hbox{$\;<$\kern-0.75em\raise-1.1ex\hbox{$\sim\;$}}}

\def\beqn#1{\begin{equation}\label{#1}}
\def\eeqn{\end{equation}}

\def\beqa#1{\begin{eqnarray}\label{#1}}
\def\eeqa{\end{eqnarray}}

\def\Z2{$\mathcal{Z_2}$}

\newcommand {\ignore}[1]{}

\begin{document}

\preprint{INR-TH-2019-023}
\preprint{CERN-TH-2019-217}

\title{Cosmological Parameters and Neutrino Masses \\
from the Final Planck and Full-Shape BOSS Data}

\author{Mikhail M. Ivanov}\email{mi1271@nyu.edu}\affiliation{Center for Cosmology and Particle Physics, Department of Physics, New York University, New York, NY 10003, USA}
\affiliation{Institute for Nuclear Research of the
Russian Academy of Sciences, \\ 
60th October Anniversary Prospect, 7a, 117312
Moscow, Russia
}
\author{Marko Simonovi\'c}\email{marko.simonovic@cern.ch}\affiliation{Theoretical Physics Department, CERN, \\ 1 Esplanade des Particules, Geneva 23, CH-1211, Switzerland
}
\author{Matias Zaldarriaga}\email{matiasz@ias.edu}\affiliation{School of Natural Sciences, Institute for Advanced Study,\\1 Einstein Drive, Princeton, NJ 08540, USA
}

\begin{abstract} 
We present a joint analysis of 
the Planck cosmic microwave background (CMB) 
and Baryon Oscillation Spectroscopic Survey (BOSS) final data releases.
A key novelty of our study is the use of a new 
full-shape (FS) likelihood for the redshift-space galaxy power spectrum of the BOSS data,
based on an improved perturbation theory template.
We show that 
the addition of the redshift space galaxy clustering measurements 
 breaks degeneracies 
present in the CMB data alone and 
tightens constraints on cosmological parameters.
Assuming the minimal $\Lambda$CDM cosmology with massive neutrinos, 
we find the following late-Universe parameters:
the Hubble constant \mbox{$H_0=67.95^{+0.66}_{-0.52}$ km s$^{-1}$Mpc$^{-1}$}, the matter density fraction 
\mbox{$\Omega_m=0.3079^{+0.0065}_{-0.0085}\,$}, the mass fluctuation amplitude 
\mbox{$\sigma_8=0.8087_{-0.0072}^{+0.012}\,$},
and an upper limit on the sum of neutrino masses \mbox{$M_{\text{tot}} <0.16\,$ eV} ($95\%$ CL).
This can be contrasted with the Planck-only measurements: \mbox{$H_0=67.14_{-0.72}^{+1.3}$} km s$^{-1}$Mpc$^{-1}$,  
$\Omega_m=0.3188^{+0.0091}_{-0.016}\,$, \mbox{$\sigma_8=0.8053_{-0.0091}^{+0.019}\,$},
and \mbox{$M_{\text{tot}} <0.26\,$ eV} ($95\%$ CL).
Our bound on the sum of neutrino masses relaxes once the hierarchy-dependent priors from the oscillation experiments are imposed.
The addition of the new FS likelihood also constrains the effective number of extra relativistic
degrees of freedom, \mbox{$N_{\text{eff}}=2.88\pm 0.17$}.
Our study shows that the current FS
and the pure baryon acoustic oscillation data 
add a similar amount of information in combination with the Planck likelihood. 
We argue that this is just a coincidence given the BOSS volume and efficiency of the current reconstruction algorithms.
In the era of future surveys FS will play a dominant role in cosmological parameter measurements.
\end{abstract}

\maketitle

\section{Introduction}

Planck cosmic microwave 
background (CMB) data have been the leading cosmological probe with unprecedented measurement
of cosmological parameters \cite{Aghanim:2018eyx}. Powerful as it is, the CMB data 
possess some internal 
parameter degeneracies, which compromise 
the accuracy of cosmological constraints, especially for
non-minimal extensions of the base $\L$CDM. 
A way to break some of these degeneracies is to use additional information form the large-scale structure (LSS) surveys.
The most well-known example is a combination of the Planck likelihood with the geometric location of the baryon acoustic oscillations (BAO) peak inferred from the galaxy correlation function \cite{Alam:2016hwk}.
There are two major reasons why this combination is often employed. 
First, the BAO peak is relatively easy to measure and it is very robust against various possible systematic effects of spectroscopic galaxy surveys. Second, the reconstruction algorithms used to ``sharpen'' the BAO feature exploit  higher $n$-point correlation functions of the nonlinear density field, which significantly improves the measurement of the location of the BAO peak \cite{Eisenstein:2006nk,Padmanabhan:2008dd,Schmittfull:2015mja,Schmittfull:2017uhh}. This allows for an accurate and robust measurement of the BAO scale, which in turn breaks degeneracies of the Planck likelihood \cite{Alam:2016hwk,Aghanim:2018eyx}. 

One important example where the BAO information plays a notable role is the constraint on the sum of neutrino masses $M_{\text{tot}}$. Significant 
efforts from both particle physics and cosmology confined this parameter
to a narrow range 
\be
\label{eq:constnow}
0.06\;\text{eV} <M_{\text{tot}} < 0.12\;\text{eV}\,,
\ee 
where the upper bound 
is a 95$\%$ confidence interval from observations of temperature and polarization
fluctuations in the CMB along with BAO in the distribution of different tracers of matter \cite{Aghanim:2018eyx},
whereas the lower limit is given by the flavor oscillations experiments\footnote{Assuming the normal hierarchy 
(the three states satisfy the hierarchy $m_1 \lesssim m_2 \ll m_3$) and that one eigenstate has a zero mass.
Note that an upper bound in Eq.~\eqref{eq:constnow} was derived without assuming the lower bound from oscillation experiments.
This point will be discussed in Sections \ref{sec:met} and \ref{sec:res}. 
} \cite{Esteban:2018azc}.
Remarkably, the combination of CMB and LSS  data 
(e.g.~\cite{Palanque-Delabrouille:2015pga,Cuesta:2015iho,Alam:2016hwk,Upadhye:2017hdl,Doux:2017tsv,Vagnozzi:2017ovm,Palanque-Delabrouille:2019iyz}, also see \cite{RoyChoudhury:2019hls} and references therein)
gives a much tighter upper bound on the total neutrino mass than
laboratory experiments like KATRIN \cite{Aker:2019uuj}.

While this and other similar examples show the importance of combining the BAO data with the CMB observations, 
the position of the BAO peak (including reconstruction) represents
only a part of cosmological
information encoded in the clustering pattern of galaxies in redshift space. 
Complementary information is in the broadband of the power spectrum (as well as higher $n$-point functions). 
This information is naturally extracted using the full-shape (FS) analysis. In this approach the whole power spectrum is exploited and, unlike  the BAO measurement alone, 
all cosmological parameters can be constrained independently of the CMB data \cite{Ivanov:2019pdj,Leonardo:2019me,Colas:2019ret}.\footnote{Similar results were later re-obtained in Ref.~\cite{Troster:2019ean} using the old BOSS pipeline.
}
Remarkably, the FS information allows one to measure the late-time matter density fraction $\Omega_{m}$ to $3\%$
and the Hubble parameter $H_0$ to $2\%$ precision without the shape and sound horizon priors from the CMB.  
These measurements are not possible with the BAO and full-shape studies that are based on scaling 
parameters (see \cite{Alam:2016hwk,Beutler:2016arn}).
The first goal of this paper is to combine the new FS likelihood of the BOSS data presented in \cite{Ivanov:2019pdj} with the Planck CMB likelihood and measure the cosmological parameters.

Having BAO and FS analyses at hand, one immediate question is to ask how do they compare. 
It is hard to give a simple and intuitive answer for several reasons.
On the one hand, the reconstructed BAO feature is sharper, 
but the broadband can be measured much better (the amplitude of the BAO wiggles is a few percent of the broadband). 
On the other hand, the broadband has no strong features and its shape is uncertain due to the nonlinear evolution
and instrumental systematic effects. 
However, the FS analysis does include the damped BAO wiggles, which still contain significant amount of information.
Given all these differences, the second goal of this paper is to answer the following simple questions:
(a) How do the cosmological parameter measurements compare between the BAO and the FS analyses of the BOSS data in combination with the Planck likelihood? 
(b) How is this comparison expected to look like for future spectroscopic surveys?

To achieve these goals we focus
on two particular well-motivated models for which the BAO or FS information is expected 
to be the most relevant: $\L$CDM with massive neutrinos and 
$\L$CDM with both massive neutrinos and extra 
relativistic degrees of freedom (parameterized by their effective number $N_{\text{eff}}$). 
These extensions of the minimal $\L$CDM model can be easily accommodated by particle physics models which feature both 
sterile and usual massive left-handed neutrinos (see \cite{Abazajian:2012ys,Adhikari:2016bei} for reviews).
Note that for other non-minimal 
models, e.g.~dynamical
dark energy, the FS power spectrum likelihood is mostly saturated 
with the distance information \cite{Ivanov:2019pdj}, and it is not expected to 
perform much better than the BAO-only likelihood.

It is worth noting that the combined analyses of the CMB and FS galaxy power spectrum 
have been already performed several times~\cite{Gil-Marin:2014baa,Beutler:2014yhv,Cuesta:2015iho,Alam:2016hwk,Grieb:2016uuo,Sanchez:2016sas,Upadhye:2017hdl}. 
These analyses were based on approximate phenomenological models 
for the non-linear power spectrum (or the correlation function). 
Even though these models capture the
main qualitative effects of the non-linear clustering 
and redshift-space distortions, their use can lead to systematic biases in the parameter inference. 
These biases may be small given the errorbars of the BOSS survey, 
but can become significant for the future high-precision LSS surveys like Euclid \cite{Amendola:2016saw}
or DESI \cite{Aghamousa:2016zmz}.
In this paper we reanalyze the 
Planck and the FS BOSS legacy data using  
the most advanced perturbation theory model that is available to date.

Our theoretical model is an improved version of the one-loop Eulerian perturbation theory, which includes corrections that parametrize the effects of complicated
short-scale physics. These corrections can be consistently taken into account within the effective 
field theory framework \cite{Baumann:2010tm,Carrasco:2012cv}.
This model is described in detail in Refs.~\cite{Senatore:2014vja,Senatore:2014eva,Perko:2016puo,Leonardo:2019me,Chudaykin:2019ock,Ivanov:2019pdj}.
The main difference with respect to previous studies is the implementation of infrared (IR) resummation
and the presence of the so-called ``counterterms.'' 
IR resummation describes the non-linear
evolution of baryon acoustic oscillations, which was independently formulated within several 
different but equivalent frameworks \cite{Senatore:2014via,Vlah:2015sea,Vlah:2015zda,Baldauf:2015xfa,Blas:2015qsi,Blas:2016sfa,Ivanov:2018gjr}.
The major novelties compared to the previous models are: (i) {The non-linear damping applies 
only to the oscillating (``wiggly'') part of the matter power spectrum, 
(ii) It does not require any fitting parameters,
and (iii) It includes corrections beyond the commonly used exponential suppression. As for the counterterms, their presence is required in order to capture the 
effects of poorly known short-scale physics \cite{Pueblas:2008uv,Baumann:2010tm,Carrasco:2012cv} on the long-wavelength fluctuations. In particular, these corrections provide an effective description of 
the baryonic feedback \cite{Lewandowski:2014rca}, 
higher derivative and velocity biases \cite{Desjacques:2016bnm},
and the redshift-space distortions \cite{Senatore:2014vja} including 
the so-called ``fingers-of-God'' effect \cite{Jackson:2008yv}. 

This paper is structured as follows. We discuss our methodology, datasets, and the treatment of massive neutrinos
in Sec.~\ref{sec:met}. Sec.~\ref{sec:res} contains main results. In Sec.~\ref{sec:wiggles_FS} we 
present a mock analysis of the simulated BOSS data that quantifies the amount of information
from the BAO and FS measurements.
Finally, Sec.~\ref{sec:concl}
draws conclusions.

\begin{table*}[t!]
  \begin{tabular}{|c||c|c|c||c|c|c|} \hline
  & \multicolumn{3}{|c|}{$\nu\L$CDM} & \multicolumn{3}{|c|}{$\nu\L$CDM~+~$N_{\text{eff}}$}\\  \hline
     \hline
    Parameter  &  Planck & {\small Planck~+~BAO }  &  {\small Planck~+~FS }   &  Planck 
     &   {\small Planck~+~BAO}
    &  {\small Planck~+~FS}
      \\ [0.2cm]
 \hline 
  $100~\omega_{b}$  &  $2.238_{-0.015}^{+0.016}$ & 
  $2.245_{-0.014}^{+0.014}$ 
  & $2.247_{-0.013}^{+0.015}$ 
  & $2.224_{-0.023}^{+0.023}$
  & $2.240_{-0.019}^{+0.019}$
  & $2.233_{-0.019}^{+0.019}$ \\ \hline
  $\omega_{cdm}$  & $0.1201_{-0.0014}^{+0.0013}$ 
  & $0.11919_{-0.00099}^{+0.00099}$ 
  & $0.11893_{-0.001}^{+0.00097}$  
  & $0.1181_{-0.0031}^{+0.003}$ 
  & $0.1182_{-0.0031}^{+0.0029}$ 
  & $0.1166_{-0.0028}^{+0.0026}$ 
   \\ \hline
  $100~\theta_{s}$   & {\small $1.04187_{-0.00030}^{+0.00030}$}
  & $1.04195_{-0.00029}^{+0.00029}$ 
  & $1.04196^{+0.00028}_{-0.00028}$
  & $1.04220_{-0.00054}^{+0.00051}$
  & $1.04210_{-0.00052}^{+0.0005}$ 
  & $1.04234_{-0.0005}^{+0.00049}$ \\ \hline
$\tau$   & $0.0543_{-0.0079}^{+0.0074}$ & 
$0.05556_{-0.0076}^{+0.007}$ 
& $0.05539_{-0.0072}^{+0.0074}$ 
& $0.05341_{-0.008}^{+0.0074}$
& $0.05516_{-0.0078}^{+0.0072}$
& $0.05409_{-0.0075}^{+0.0073}$  \\ \hline
$\ln(10^{10}A_s)$   & $3.045_{-0.016}^{+0.014}$ 
& $3.045_{-0.015}^{+0.014}$ 
& $3.044_{-0.014}^{+0.014}$
& $3.037_{-0.018}^{+0.018}$
& $3.042_{-0.017}^{+0.017}$ 
& $3.035_{-0.017}^{+0.016}$  \\ 
\hline
$n_s$  
&  $0.9646_{-0.0045}^{+0.0045}$
& $0.9669_{-0.0039}^{+0.0039}$ 
& $0.967_{-0.004}^{+0.0038}$  
&$0.9588_{-0.0087}^{+0.0087}$ 
& $0.9647_{-0.0074}^{+0.0073}$& 
$0.9608_{-0.0072}^{+0.0074}$ \\ 
\hline
$M_{\text{tot}}$   & $<0.26~$ & $<0.12~$& $<0.16~$ & $<0.27$& $<0.12$ & $<0.16$  \\   \hline
$N_{\text{eff}}$   & \multicolumn{3}{|c||}{fixed $3.046$}   
& $2.90_{-0.19}^{+0.19}$ 
& $2.99_{-0.17}^{+0.17}$ 
& $2.88_{-0.17}^{+0.17}$ \\   
\hline\hline
$\Omega_m$   & $0.3188^{+0.0091}_{-0.016}$ 
& $0.3078^{+0.0060}_{-0.0071}$ 
& $0.3079^{+0.0065}_{-0.0085}$ 
& $0.324^{+0.011}_{-0.019}$ 
& $0.3090^{+0.007}_{-0.0076}$ & 
$0.3127^{+0.0080}_{-0.0091}$  \\ \hline
$H_0$   & $67.14_{-0.72}^{+1.3}$ 
& $67.97_{-0.49}^{+0.56}$ 
& $67.95_{-0.52}^{+0.66}$ 
& $66.1_{-1.6}^{+1.9}$ 
& $67.6_{-1.2}^{+1.2}$ 
& $66.8_{-1.2}^{+1.2}$  \\ \hline
$\sigma_8$   & $0.8053_{-0.0091}^{+0.019}$ 
&$0.8135_{-0.0073}^{+0.01}$ 
& $0.8087_{-0.0072}^{+0.012}$
& $0.798_{-0.013}^{+0.022}$
& $0.811_{-0.011}^{+0.012}$ 
& $0.8015_{-0.011}^{+0.013}$ \\ \hline
\end{tabular}
\caption{Mean values and 68\% CL minimum credible
intervals for the parameters of the $\nu\Lambda$CDM (left three columns) and 
$\nu\Lambda$CDM$~+~N_{\text{eff}}$ (right three columns) models as extracted 
from the Planck, Planck~+~BAO, and Planck~+~FS data, 
presented as ``mean$^{+1\sigma}_{-1\sigma}$.''
For $M_{\text{tot}}$ we quote the 95\% CL upper limit in units of eV. $H_0$
is quoted in km/s/Mpc.
}
\label{table0}
\end{table*}

\section{Methodology}
\label{sec:met}

In our main analysis the Markov-Chain Monte Carlo (MCMC) chains sample seven cosmological parameters of the minimal $\L$CDM
with massive neutrinos ($\omega_{b}$,$\omega_{cdm}$,$\theta_s$,$A_s$,$\tau$,$n_s$,$M_{\text{tot}}$),
where $\omega_{b}$, $\omega_{cdm}$ are physical densities of baryons and dark matter respectively,
$\t_s$ is the angular acoustic scale of the CMB,
$A_s$ and $n_s$ are the amplitude and the tilt of the primordial spectrum of scalar
fluctuations, $\tau$ denotes the reionization optical depth,
and $M_{\text{tot}}$ is the sum of neutrino mass eigenstates.
Additionally, we
run an analysis with varied $N_{\text{eff}}$ which was fixed to the standard value $3.046$
in the baseline run. Throughout this paper 
we approximate the neutrino sector with three 
degenerate massive states. 
This approximation is very accurate both for the current and future surveys. 
The difference between the exact mass splittings and 
and the degenerate state approximation is negligible once the proper lower priors are imposed \cite{Lesgourgues:2006nd,DiValentino:2016foa,Vagnozzi:2017ovm,RoyChoudhury:2019hls}.

From the Planck side, we use the baseline TTTEEE~+~low $\ell$~+~lensing likelihood from the 2018 data release \cite{Aghanim:2018eyx} as implemented in \texttt{Montepython v3.0} \cite{Brinckmann:2018cvx}, see Ref.~\cite{Aghanim:2019ame}
for likelihood details. 
In addition to the cosmological parameters, we also vary 21 nuisance parameters that describe 
foregrounds, beam leakage, and other instrumental effects \cite{Aghanim:2019ame}.
One difference with respect to the baseline Planck analysis is that we model the 
non-linear corrections to the CMB lensing potential with one-loop perturbation theory. 
The reason is that the one-loop power spectrum captures the behavior  
of the matter power spectrum on mildly non-linear scales
much better than the commonly used fitting formulas like HALOFIT.
Strictly speaking, the one-loop power spectrum cannot be applied to very non-linear
scales.
However, for $\L$CDM the one-loop power spectrum matches the HALOFIT formula with $\sim 20\%$
accuracy down to $k\sim 1~h$Mpc$^{-1}$. 
Moreover, the one-loop expression is more reliable for non-minimal cosmological models, for which the HALOFIT was simply not calibrated.  
We have run the Planck baseline analysis both with the HALOFIT and one-loop perturbation theory
and found identical results.

To quantify the constraining power of the BOSS FS likelihood, 
we compare our results with the 
joint analysis of the Planck and the consensus BAO measurements
based on the same BOSS data \cite{Alam:2016hwk}. 
Note that this BAO likelihood is somewhat less constraining 
compared to the one used by the Planck collaboration \cite{Aghanim:2018eyx}, which also included e.g.~data
from Ly-$\alpha$ forest absorption lines \cite{Bourboux:2017cbm} and quasar clustering \cite{Ata:2017dya}.
The consensus BAO measurements of BOSS were obtained by the so-called
density field reconstruction \cite{Eisenstein:2006nk,Padmanabhan:2008dd}, 
which sharpens the BAO feature but distorts the broadband shape,
which is then marginalized over.\footnote{It is worth mentioning that a promising way to extract cosmological information from galaxy catalogs can be a consistent reconstruction of the full initial density field beyond the BAO~\cite{Feng:2018for,Modi:2018cfi,Schmittfull:2018yuk,Schmidt:2018bkr,Elsner:2019rql}.} 

The main analysis of this paper will be based on the full-shape galaxy power spectrum likelihood from the BOSS data release 12 (year 2016) \cite{Alam:2016hwk}, which includes 
the monopole and quadrupole moments at wavenumbers up to \mbox{$k_{\text{max}}=0.25\,h$/Mpc}.
Details of this likelihood can be found in Ref.~\cite{Alam:2016hwk}. 
The BOSS DR12 includes four independent 
datasets corresponding to different galaxy populations observed
across two non-overlapping redshift bins with $z_{\text{eff}}=0.38$ and $z_{\text{eff}}=0.61$.
For each dataset we use 7 nuisance parameters to describe galaxy bias, baryonic feedback, ``fingers-of-God''
and other effects of poorly known short-scale physics, which totals to 28 additional 
free parameters in the joint BOSS FS likelihood. Our methodology for the BOSS full-shape analysis is identical to the one used in Ref.~\cite{Ivanov:2019pdj},
where one can find further details of the theoretical model, covariance matrix, and the window function treatment. 
Additionally, in this work we account for fiber collisions by implementing 
the effective window method \cite{Hahn:2016kiy}. 
In agreement with previous works \cite{Ivanov:2019pdj,Leonardo:2019me}, 
we have found that the effect of fiber collisions is 
largely absorbed into the nuisance parameters and
has negligible impact on the estimated cosmological parameters.
In the present analysis we ignore any correlation between 
the BOSS and the CMB data.
The cross-correlation of BOSS galaxies with the CMB temperature has not yet been detected \cite{Nicola:2016eua}, while the correlation with the CMB lensing is small on the mildly nonlinear scales 
\cite{Doux:2017tsv}. 
Thus, treating the BOSS and Planck data as independent is a reasonable 
approximation given 
the current errorbars.

The presence of massive neutrinos requires a modification of the standard perturbative approach
to galaxy clustering.
Neutrino free-streaming makes the growth of matter fluctuations scale-dependent, which invalidates the common perturbative schemes that are based 
on the factorization of time evolution in the perturbation theory kernels 
(the so-called Einstein-de Sitter approximation). 
A fully consistent description requires a proper calculation of scale-dependent Green's functions, 
see Refs.~\cite{Blas:2014hya,Senatore:2017hyk}. 
However, this description is quite laborious and has not yet been extended to galaxies in
redshift space. Given the errorbars of the BOSS survey, 
one may consider the effect of massive neutrinos perturbatively and employ some approximations.
In particular, we will use standard expressions for the one-loop integrals 
computed in the Einstein-de Sitter (EdS) approximation, 
but with the exact linear power spectrum obtained in the presence of massive neutrinos \cite{Takada:2005si}.
For calculations based on a two-fluid extension of standard perturbation theory, 
this prescription has been checked to agree with the 
full treatment up to a few percent difference \cite{Blas:2014hya}.
This result was recently confirmed in effective field theory in Ref.~\cite{Senatore:2017hyk}.
This work showed that the leading effects of non-linear neutrino backreaction is 
captured by the counterterms, 
which also absorb the difference between proper Green's functions
and the EdS approximation on large scales.
We will also employ the ``cb'' prescription, i.e. assume that
galaxies trace only dark matter and baryons, and not the total 
matter density that includes the massive neutrinos. 
This prescription was advocated on the basis of N-body simulations in Refs.~\cite{Villaescusa-Navarro:2013pva,Castorina:2013wga,Costanzi:2013bha,Castorina:2013wga,Castorina:2015bma,Villaescusa-Navarro:2017mfx}.
Furthermore, Refs.~\cite{Raccanelli:2017kht,Vagnozzi:2018pwo} pointed out its importance for parameter
inference.
Following the ``cb'' prescription, 
we evaluate the loop integrals 
using the standard perturbation theory redshift-space kernels with the logarithmic growth rate computed only
for the baryon and dark matter components. 
The ``cb'' prescription ensures that 
the galaxy power spectrum matches N-body simulations
on large scales~\cite{Villaescusa-Navarro:2017mfx}, 
where it approaches the Kaiser prediction \cite{Kaiser:1987qv} evaluated with 
the linear bias and logarithmic growth factor $f$ for the baryon~+~dark matter fluid.

Before closing this Section it is worth mentioning that Refs.~\cite{Chiang:2017vuk,Chiang:2018laa}
found additional scale-dependence of galaxy bias even if it is defined with respect to CDM+baryons. 
It was argued that this effect is numerically very small for standard cosmology, 
but it depends linearly on the neutrino density fraction just like the other effects relevant for galaxy clustering.
We leave the impact of the neutrino-induced bias 
on cosmological parameter measurements for future study.

\begin{figure*}[ht]
\begin{center}
\includegraphics[width=1\textwidth]{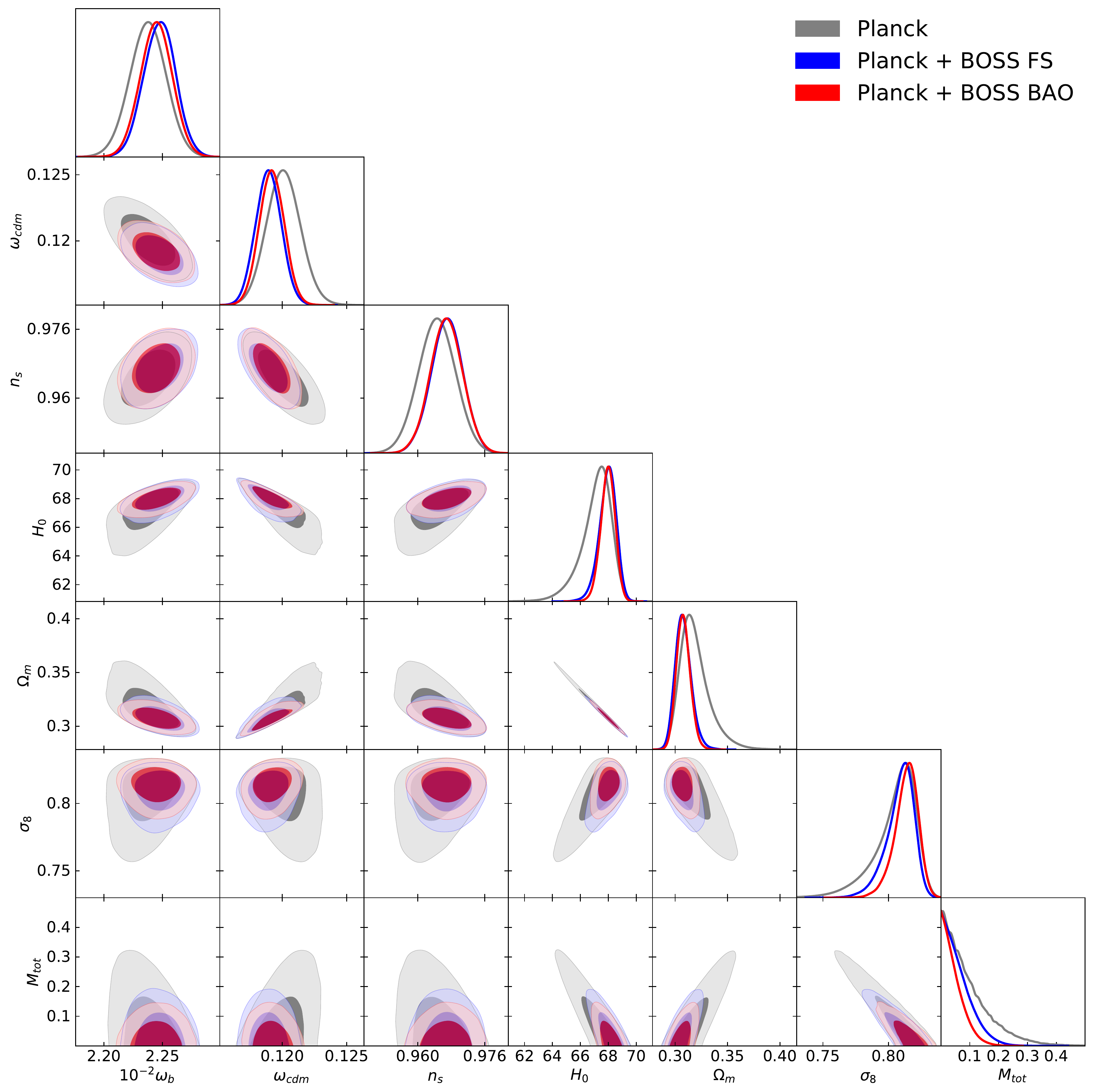}
\end{center}
\caption{
Marginalized one-dimensional posterior
  distribution and two-dimensional probability contours (at the 68\%
  and 95\% CL) for the 
parameters of the $\Lambda$CDM model with varied neutrino masses. $N_{\text{eff}}$ is fixed 
to the standard model value $3.046$. $H_0$
is quoted in km/s/Mpc, $M_{\text{tot}}$ is quoted in eV.
\label{fig:mnu} } 
\end{figure*}

\begin{figure*}[ht]
\begin{center}
\includegraphics[width=1\textwidth]{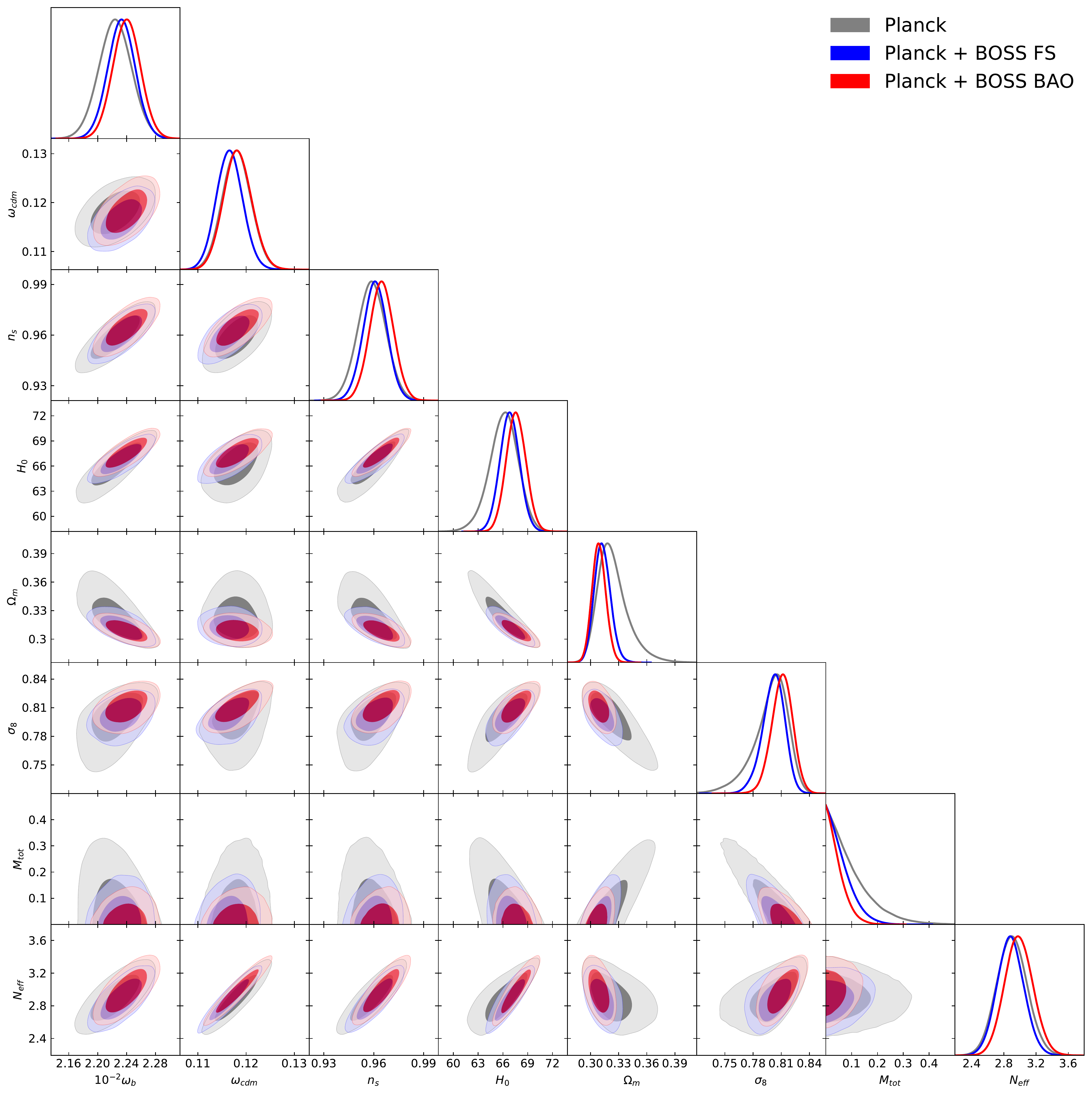}
\end{center}
\caption{
Marginalized one-dimensional posterior
  distribution and two-dimensional probability contours (at the 68\%
  and 95\% CL) for the 
parameters of the $\Lambda$CDM model with both varied neutrino masses
and the effective number of relativistic degrees of freedom $N_{\text{eff}}$.
$H_0$
is quoted in km/s/Mpc, $M_{\text{tot}}$ is quoted in eV.
\label{fig:neff} 
} 
\end{figure*}

\section{Results}
\label{sec:res}

The triangle plots with posterior densities and marginalized 
distributions for cosmological 
parameters of the $\nu\L$CDM model ($\omega_b,\omega_{cdm},n_s,H_0,\Omega_m,\sigma_8,M_{\text{tot}}$) 
are shown in Fig.~\ref{fig:mnu}. 
A similar plot obtained for a model with free 
$N_{\text{eff}}$ (dubbed as $\nu\L$CDM~+~$N_{\text{eff}}$) is displayed in Fig.~\ref{fig:neff}. 
For comparison, we also show the contours obtained by analyzing the Planck 
data only. Results of this analysis are in good agreement with the ones 
reported by the Planck collaboration~\cite{Aghanim:2018eyx}.\footnote{\href{https://wiki.cosmos.esa.int/planck-legacy-archive/images/b/be/Baseline_params_table_2018_68pc.pdf}{
\textcolor{blue}{https://wiki.cosmos.esa.int/planck-legacy-archive/images/b/be/Baseline$\_$params$\_$table$\_$2018$\_$68pc.pdf}} 
} 
The marginalized limits are presented in Table~\ref{table0}.

The BOSS data notably improve the limits on the late-time parameters $H_0$,
$\Omega_m$, $\sigma_8$, and the neutrino mass $M_{\text{tot}}$.
This happens mainly due to the breaking of degeneracies 
between $H_0$ and other cosmological parameters in the CMB likelihood.
This is not surprising, as the precision 
of $H_0$ measurement from the BOSS FS data alone rivals that of the CMB.
The main improvement on the sum of neutrino
masses brought by the BOSS data also comes from a better $H_0$ determination (this result was foreseen long ago
in Ref.~\cite{Hamann:2010pw}). 
$H_0$ and $M_{\text{tot}}$ are anticorrelated in the CMB data, and the BOSS likelihood pulls $H_0$ to slightly higher values \cite{Ivanov:2019pdj}, which
pushes the neutrino masses closer to the origin. 
However, the BOSS data at the same time prefer a somewhat low value of $\sigma_8$ \cite{Ivanov:2019pdj} which pulls the neutrino masses in the opposite direction. 
This is reflected in our $95\%$ CL limit
\mbox{$0.16$ eV}, which is higher than the Planck~+~BAO measurement 
\mbox{$0.12$ eV}. 
We stress that this relaxation does not imply that the FS data has less statistical power than the BAO. 
On the contrary, it is a result of taking into account new information that the BOSS clustering amplitude is lower than 
the Plank~+~BAO prediction.


It is instructive to see how much the neutrino mass bounds depend on the priors.
Following the Planck methodology, we have imposed an unphysical zero lower limit in our baseline analysis.
However, the physical priors corresponding to the normal or inverted hierarchies
(NH and IH in what follows, respectively) can 
notably change the result. To estimate this effect we have 
resampled our chains with the physical lower priors ($0.06$~eV for DH and $0.1$~eV for IH) and obtained the following bounds (at 95$\%$CL):
\be 
\begin{split}
& M_{\text{tot}}<0.18\,\text{eV}\quad (\text{NH, Planck~+~FS})\,,\\
&M_{\text{tot}}<0.21\,\text{eV}\quad (\text{IH, Planck~+~FS})\,.\\
\end{split}
\ee
These values can be compared with the Planck~+~BAO results which were extracted from our chains
by a similar resampling,
\be 
\begin{split}
& M_{\text{tot}}<0.15\,\text{eV}\quad (\text{NH, Planck~+~BAO})\,,\\
&M_{\text{tot}}<0.18\,\text{eV}\quad (\text{IH, Planck~+~BAO})\,.\\
\end{split}
\ee

As far as the base cosmological parameters are concerned, 
the improvement from the FS data (which embody the unreconstructed BAO)
for $\nu\L$CDM
is comparable to that from the reconstructed BAO measurements \cite{Aghanim:2018eyx}. 
This reflects the fact that the shape of the matter power spectrum does not
contribute significantly to the cosmological constraints on 
the physical densities of baryons and dark matter, which are dominated
by Planck.

One may expect that the shape information can be more important  
is the model with additional relativistic degrees of freedom. 
However, in this model the CMB degeneracy direction in the plane $\omega_{cdm}-H_0$
changes its orientation compared to the base $\L$CDM and accidentally becomes aligned with the 
degeneracy direction of the BOSS data. 
Due to this coincidence the parameter degeneracies from the two datasets do not get broken, and the 
improvement
from their combination is quite modest. 
Importantly, the posterior contour in the $\omega_{cdm}-H_0$ plane is shifted 
down as a consequence of 
the preference of the BOSS data for low $\omega_{cdm}$ \cite{Ivanov:2019pdj}.
This also produces some $\sim 0.5\sigma$ 
shifts in cosmological parameters as compared 
to the Planck~+~BAO analysis.
In particular, we find 
\be 
 \begin{aligned}
N_{\text{eff}}&=2.88\pm 0.17 \\
H_0&=66.8\pm 1.2
\end{aligned}
\Bigg\} \quad 
 \begin{aligned}
 \text{(68 $\%$, Planck~+~FS),}\\
 \end{aligned}
\ee
which can be contrasted with  
\be 
 \begin{aligned}
N_{\text{eff}}&=2.99\pm 0.17 \\
 H_0&=67.6\pm 1.2
\end{aligned}
\Bigg\} \quad 
 \begin{aligned}
 \text{(68 $\%$, Planck~+~BAO),}\\
 \end{aligned}
\ee
where $H_0$ is quoted in km/s/Mpc.
Note that the FS and the BAO data pull the mean of $N_{\text{eff}}$ in different directions. 
Moreover, unlike the FS data, 
the BAO notably shift the means of other parameters, e.g.~$\omega_{cdm}$ and $H_0$.
This shows that the full-shape and BAO data: 
(i) Contain different information, 
(ii) Have similar statistical 
powers in combination with Planck. 
The interpretation of these results is that most of the improvement 
in the joint constraint comes from breaking of geometric degeneracy between $H_0$
and other cosmological parameters. Both the BAO and FS have the same amount of 
geometric information that primarily constrains $H_0$ for the models that we considered \cite{Ivanov:2019pdj},
and hence the errorbars are very similar. 
However, the additional shape and clustering amplitude information contained 
in the FS data are not negligible, 
and their addition leads to $\sim 0.5\sigma$ shifts of the Planck~+~FS posteriors compared to Planck~+~BAO.

Finally, let us remark that we varied $N_{\text{eff}}$ together with the neutrino mass, 
but this choice does not degrade our limits compared to a fit with 
fixed $M_{\text{tot}}$. 
The reason is that  
the Planck data itself clearly distinguish between the two effects because
the errorbars from the joint $M_{\text{tot}}~+~N_{\text{eff}}$ fit 
are the same as in the individual $M_{\text{tot}}$ and $N_{\text{eff}}$ 
runs \cite{Aghanim:2018eyx}. 
This is also true for the BOSS likelihood, as we have obtained identical constraints on 
$M_{\text{tot}}$ that are twice stronger than Planck
both with fixed and varied $N_{\text{eff}}$. This
suggests that the two effects are 
clearly discriminated by the BOSS data too.

\begin{figure*}[ht]
\begin{center}
\includegraphics[width=0.7\textwidth]{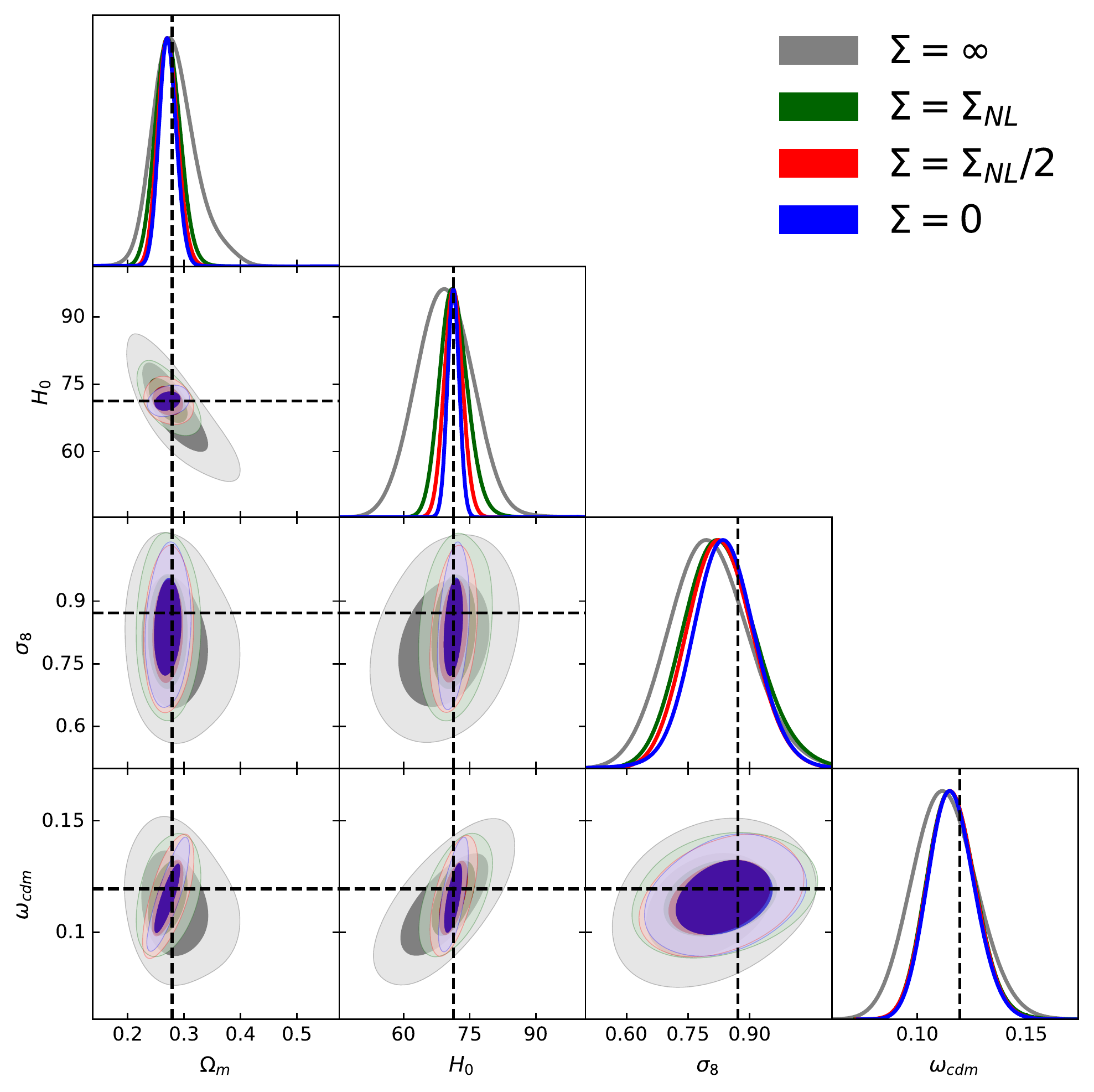}
\end{center}
\caption{
Marginalized one-dimensional posterior
  distribution and two-dimensional probability contours (at the 68\%
  and 95\% CL) for the 
parameters of simulated mock BOSS data. 
The shown are four cases corresponding to different amount of the BAO smoothing, see the main text
for further details.
\label{fig:mock} 
} 
\end{figure*}

\section{Wiggles vs.~Broadband}
\label{sec:wiggles_FS}

In the previous Section we have presented results for the two different analyses, Planck~+~BAO and Planck~+~FS. 
As argued in Introduction, the information extracted form the galaxy clustering in these two methods is quite different, 
yet the error bars in the two analyses are identical.\footnote{As discussed in the previous section, the 95$\%$ upper 
bound on $M_{\rm tot}$ in the FS analysis is larger. However, this is due to low $\sigma_8$ measured by BOSS compared to 
the CMB, which pulls the upper bound to the larger value. This does not happen in the BAO analysis, since $\sigma_8$ is not measured.} 
This is a very striking feature of our results and it requires an explanation. 
In this section we investigate the information content in the BAO and FS analyses in detail and show that the identical error 
bars are just a coincidence for the given volume of the BOSS survey and the given BAO reconstruction efficiency. We will argue that for larger future spectroscopic 
surveys the FS analysis will eventually be more powerful in constraining the cosmological parameters.

In order to compare the amount of information in the reconstructed BAO wiggles with the amount of information in 
the full shape power spectrum (which embeds the unreconstructed BAO),
we analyzed several sets of the simulated mock data which mimic the actual BOSS sample, but 
have different amplitudes of the BAO wiggles. 
This exercise is analogous to that performed in Ref.~\cite{Ivanov:2019pdj},
where one can find further details of our mock dataset. 
We generated four sets of power spectra multipoles
for the low-z ($z_{\text{eff}}=0.38$) DR12 North Galactic Cap (NGC) sample
with a different amount of the BAO damping and
analyzed them using the same pipeline appropriately modified in each case. 
The mock data were assigned the covariance of the real sample.
For clarity, we analyze the mock BOSS data \textit{per se}, i.e.~without the Planck likelihood.\footnote{There are two effects that contribute to the constraints in combination with Planck. 
First, it is the size of the errorbar itself and second, 
it is the orientation of the posterior contours (e.g. $H_0-\omega_{cdm}$) w.r.t. the Planck ones, 
which is different for BAO and FS. 
}

To understand our method, recall that the BAO damping at leading order can be described as 
\be
\label{eq:irres}
P_{\text{IR res, LO}}(k,\mu)=P_{\text{nw}} + \e^{-\Sigma^2 k^2}P_{\text{w}}\,,
\ee
where $P_{\text{nw}}$ and $P_{\text{w}}$ are the de-wiggled broadband and wiggly parts of the linear power spectrum respectively.
We also introduced $\mu\equiv \cos({\bf z},\k)$, where ${\bf z}$ is the line-of-sight vector.
The theoretical prediction for the damping factor $\Sigma$ is given by \cite{Baldauf:2015xfa,Blas:2016sfa,Ivanov:2018gjr},
\be
\label{eq:sigma}
\begin{split}
 &\Sigma^2 = \Sigma^2_{\text{NL}}\equiv (1+f\mu^2(2+f))\\
 &\times \frac{4\pi}{3}\int_0^{k_S}dqP_{\text{nw}}(q)\left[1-j_0\left(qr_d\right)+2j_2\left(qr_d\right)\right]\,,
\end{split}
\ee
where $f$ is the logarithmic growth factor, $j_\ell(x)$ are spherical Bessel functions, $k_S$ is an arbitrary scale 
separating the resummed soft modes and $r_d$ is the sound horizon at the drag epoch. 
We emphasize that the leading order expression \eqref{eq:irres} has a non-negligible dependence on $k_S$, which greatly 
reduces after computing the one-loop correction to Eq.~\eqref{eq:irres}~\cite{Blas:2016sfa}. 
Note that we use Eq.~\eqref{eq:irres} only for illustration purposes.
In the actual analysis we compute the full one-loop IR resummed 
expression with appropriately modified values of~$\Sigma$.

The four mock samples are characterized by four different BAO damping factors
$\Sigma=\infty, \Sigma_{\text{NL}}, \Sigma_{\text{NL}}/2, 0$, where $\Sigma_{\text{NL}}$ is the theoretically 
predicted amount of the BAO damping \eqref{eq:sigma}.
The first case corresponds to the pure broadband information without any wiggles.
The second case mimics the real physical situation 
and reproduces the actual constraints from the FS analysis of the BOSS low-z NGC data sample. 
The third situation corresponds to 
the combination of the broadband with the standard BAO reconstruction, which reduces the damping
by a factor of two~\cite{Beutler:2016ixs}.
Finally, the fourth scenario features the full BAO wiggles, which are not affected by the non-linear smearing.
This case corresponds to the joint analysis the broadband~+~optimally reconstructed BAO wiggles, 
which is the best case scenario for BAO reconstruction.\footnote{The standard reconstruction technique does not 
fully restore the linear amplitude of the BAO wiggles \cite{Beutler:2016ixs}. 
However, more sophisticated methods have potential to achieve almost optimal efficiency. 
One example is the iterative reconstruction,
so far applied only to dark matter in real space~\cite{Schmittfull:2017uhh}. 
Another example is the neural network-based algorithm of Ref.~\cite{Modi:2018cfi}, 
which is close to optimal for halos.
It will be interesting to see how much these more advanced 
approaches can improve BAO reconstruction in the realistic case
of biased tracers in redshift space.}

\begin{table}[t!]
  \begin{tabular}{|c|c|c|c|c|} \hline
    Param.  & \;$\Sigma = \infty$\; & \; $\Sigma = \Sigma_{\text{NL}}$ \; & \; $\Sigma =\Sigma_{\text{NL}}/2$\; & \; $\Sigma = 0$\; \\ [0.2cm]
 \hline 
  $\omega_{cdm}$  & $0.112_{-0.015}^{+0.014}$ & $0.116_{-0.011}^{+0.010}$  & $0.116_{-0.011}^{+0.010}$  & $0.116_{-0.011}^{+0.010}$ \\ \hline
  $H_0$   & $69.3^{+6.3}_{-6.1}$ & $71.4^{+2.9}_{-3.4}$ & $71.3^{+1.9}_{-2.1}$  & $71.3^{+1.2}_{-1.4}$\\ \hline
  $\sigma_8$   & $0.802_{-0.100}^{+0.091}$ & $0.828_{-0.092}^{+0.082}$ & $0.828_{-0.075}^{+0.076}$ & $0.837_{-0.071}^{+0.072}$\\ \hline \hline
$\Omega_m$   & $0.284^{+0.031}_{-0.074}$ & $0.271^{+0.021}_{-0.021}$ & $0.271^{+0.017}_{-0.018}$ & $0.271^{+0.015}_{-0.016}$\\ \hline
\end{tabular}
\caption{Mean values and 68\% CL minimum credible
intervals for the parameters extracted from the simulated data mocking the BOSS DR12 low-z NGC sample.
The values of $H_0$
are quoted in units of km/s/Mpc.
}
\label{table:mock}
\end{table}

For the purposes of this Section, we focus on the following set of cosmological parameters ($H_0,\omega_{cdm}$,$\sigma_8$)
and use the Planck priors for $\omega_b$ and $n_s$. We fixed $M_{\text{tot}}=0$ in our simulated data.
The chosen fitting parameters represent three different sources of information encoded in the power-spectrum multipoles:
geometric distance ($H_0$), shape of the transfer functions ($\omega_{cdm}$), and redshift-space distortions ($\sigma_8$)\footnote{In 
$\L$CDM the logarithmic growth rate $f$ is fixed by $\Omega_m$ and $H_0$ (modulo a small effect due to massive neutrinos), 
which are extracted from the monopole.}.
The results of our analysis are displayed in Fig.~\ref{fig:mock} and Table~\ref{table:mock}. We also show the derived parameter
$\Omega_m$, which comes from a combination of the shape and distance information.

The relative importance of the BAO wiggles compared to the broadband can be assessed comparing the results for the four different mock data sets. 
The first observation is that the BAO wiggles significantly affect only the $H_0$ measurement.
There is no improvement in $\omega_{cdm}$ between the reconstructed and unreconstructed cases,
and a very slight errorbar reduction for the clustering amplitude $\sigma_8$.
The second observation is that the $H_0$ constraint improves by $\sim 40\%$ (i.e.~the error reduces by $\sim \sqrt{2}$) 
in $\Sigma =\Sigma_{\text{NL}}/2$ compared to the $\Sigma =\Sigma_{\text{NL}}$ case.
Thus, we can conclude that the reconstructed high-$k$ BAO wiggles measure $H_0$ with the similar precision as the FS data. This is precisely related to 
our result that the BAO and FS have a similar amount of geometric information. 
However, this is just a coincidence. Even small modifications in the setup can change the conclusions drastically.
For example, in the case of the ideal BAO reconstruction the error 
on $H_0$ is smaller by more than a factor of $2$ compared to the standard FS analysis. This result suggests that 
$H_0$ errorbars are very sensitive to the efficiency of BAO reconstruction. While $\Sigma=0$ limit is probably impossible to get in practice, 
any improvement in the reconstruction algorithm can potentially be very important for the BOSS data analysis.

Another relevant parameter in this discussion is the volume of the survey.
Smaller statistical errors can significantly improve the cosmological constraints 
thanks to the degeneracy breaking among many nuisance parameters needed to describe the broadband.\footnote{In such cases the improvement can be much better than naive estimates using the mode counting.}
Furthermore, larger surveys include higher redshifts, where the the BAO peak is much less damped.
Thus, the expectation is that for large enough volumes, the FS should eventually win over the BAO-only analysis. 

This is indeed the case. A similar mock analysis for a Euclid-like survey \cite{Chudaykin:2019ock} 
(whose volume is roughly 10 times larger than BOSS)
has shown that even in the ideal case of the $100\%$-efficient
BAO reconstruction the errorbar on $H_0$ improves only by $\lesssim 30\%$ compared to the FS constraints, 
which should be contrasted with the $\sim 100\%$ improvement for the BOSS volume. 
Repeating the analysis as Ref.~\cite{Chudaykin:2019ock}  
for a more realistic case of $50\%$-efficient reconstruction we found the improvement 
for the Euclid data to be marginal ($\lesssim 10\%$),
which can be contrasted with the $\sim 40\%$ gain for the BOSS volume.\footnote{Inclusion of the higher order $n$-point functions
further strengthens the case for the full-shape analysis. 
For instance, Ref.~\cite{Chudaykin:2019ock} has shown that the combination of the one-loop power spectrum
and tree-level bispectrum monopole can lead to better constraints than the best case 
power spectrum analysis with the optimally reconstructed BAO wiggles. 
One may expect even more benefit from addition of the higher multipole moments of the bispectrum~\cite{Yankelevich:2018uaz}.}
These results are not very surprising, and similar trends have been already seen in several other forecasts (see for instance BAO and broadband comparison in~\cite{Aghamousa:2016zmz}).

In conclusion, comparing the amount of information in the BAO and FS analyses in detail, we find that the 
similarity between the two in combination with Planck is just a coincidence of the BOSS survey volume and efficiency of the current reconstruction algorithms. 
Our analysis suggests two main conclusions: (a) Better reconstruction algorithms or optimal combination of the FS and BAO analyses can lead to tighter constraints on cosmological parameters using the same BOSS data, 
(b) The full-shape power spectrum data will supersede the BAO measurements in the era of future galaxy surveys,
even for in the case of ideal BAO reconstruction.

\section{Conclusions}
\label{sec:concl}

We have presented a joint analysis of the final Planck CMB and BOSS galaxy power 
spectrum data. Our main results include new limits on the parameters of the minimal $\Lambda$CDM, neutrino masses, and the number of effective 
relativistic degrees of freedom.
The key new feature of our work is the use of a new BOSS full-shape power spectrum likelihood,
which is based on an improved perturbation theory model. 
This model consistently accounts for 
nonlinearities of the underlying dark matter fluid,
galaxy bias, redshift-space distortions,
and nonlinear effects of large-scale bulk flows.

We showed that the addition of the BOSS FS 
data improves the Planck-only constraints. 
The results for the minimal $\L$CDM with varied 
$M_{\text{tot}}$ are very similar to the standard Planck~+~BAO analysis. 
For the model with additional relativistic degrees of freedom 
the FS and BAO data yield comparable statistical improvement but shift the posterior in 
different directions. 
We argued that this is the effect of the additional full-shape information beyond the geometric location
of the BAO.

When combined with Planck, 
the cosmological information in the shape of the BOSS galaxy power spectrum turned out to be 
comparable to the pure geometric information extracted form the reconstructed BAO peak
for the cosmological models considered in this paper. 
However, the FS measurement will become more 
powerful than the BAO in the era of future galaxy surveys even for constraining vanilla cosmological scenarios.
Importantly, the precision of the shape parameter measurements from these surveys will be comparable 
to that of Planck, and the combination of the two will reduce the errorbars by a factor of few due 
to degeneracy breaking \cite{Chudaykin:2019ock}.
This effect will be essential for the future neutrino
mass measurements \cite{Audren:2012vy,
Font-Ribera:2013rwa,
Aghamousa:2016zmz,Archidiacono:2016lnv,Sprenger:2018tdb,Brinckmann:2018owf,Chudaykin:2019ock}. 
The presented constraints set a reference mark for future LSS and CMB observations that will
surpass Planck and BOSS.

\vspace{1cm}
\section*{Acknowledgments}

We are indebted to Anton Chudaykin for his collaboration on the initial stages of this projects. 
We are grateful to Uro$\check{\text{s}}$ Seljak for his comments on the draft of this paper. 
We thank Thejs Brinckmann, Emanuele Castorina, Antony Lewis, and Inar Timiryasov for valuable discussions.
All numerical analyses of this work were performed on the 
Helios cluster at the Institute for Advanced Study.  
MI is partially supported by the Simons Foundation's Origins of the Universe program.
MZ is supported by NSF grants AST-1409709, PHY-1521097 and PHY-1820775 the Canadian 
Institute for Advanced Research (CIFAR) Program on Gravity and the Extreme 
Universe and the Simons Foundation Modern Inflationary Cosmology initiative.

Parameter estimates presented in this paper are obtained with a modified version of the \texttt{CLASS} 
code \cite{Blas:2011rf} interfaced with the \texttt{Montepython} MCMC sampler \cite{Audren:2012wb,Brinckmann:2018cvx}. 
Perturbation theory integrals are evaluated using the method described in \cite{Simonovic:2017mhp}.
The plots with posterior densities and marginalized limits are generated with the latest version of the \texttt{getdist} package\footnote{\href{https://getdist.readthedocs.io/en/latest/}{
\textcolor{blue}{https://getdist.readthedocs.io/en/latest/}}
},
which is part of the \texttt{CosmoMC} code \cite{Lewis:2002ah,Lewis:2013hha}.


\bibliography{short.bib}

\end{document}